\documentclass[aps,prb,reprint,superscriptaddress,showpacs]{revtex4-1}

\usepackage{graphicx}% Include figure files

\begin{document}

% Use the \preprint command to place your local institutional report
% number in the upper righthand corner of the title page in preprint mode.
% Multiple \preprint commands are allowed.
% Use the 'preprintnumbers' class option to override journal defaults
% to display numbers if necessary
%\preprint{}

%Title of paper
\title{Specific heat in high magnetic fields and magnetic phase diagram of CePt$_2$In$_7$}

% repeat the \author .. \affiliation  etc. as needed
% \email, \thanks, \homepage, \altaffiliation all apply to the current
% author. Explanatory text should go in the []'s, actual e-mail
% address or url should go in the {}'s for \email and \homepage.
% Please use the appropriate macro foreach each type of information

% \affiliation command applies to all authors since the last
% \affiliation command. The \affiliation command should follow the
% other information
% \affiliation can be followed by \email, \homepage, \thanks as well.
\author{Y. Krupko}
\author{A. Demuer}
\affiliation{Laboratoire National des Champs Magn\'{e}tiques Intenses (LNCMI-EMFL), CNRS, UJF, 38042 Grenoble, France}
\author{S. Ota}
\affiliation{Graduate School of Science and Technology, Niigata University, Niigata 950-2181, Japan}
\author{Y. Hirose}
\author{R. Settai}
\affiliation{Department of Physics, Niigata University, Niigata 950-2181, Japan}
\author{I. Sheikin}
\email[]{ilya.sheikin@lncmi.cnrs.fr}
%\homepage[]{Your web page}
%\thanks{}
%\altaffiliation{}
\affiliation{Laboratoire National des Champs Magn\'{e}tiques Intenses (LNCMI-EMFL), CNRS, UJF, 38042 Grenoble, France}

%Collaboration name if desired (requires use of superscriptaddress
%option in \documentclass). \noaffiliation is required (may also be
%used with the \author command).
%\collaboration can be followed by \email, \homepage, \thanks as well.
%\collaboration{}
%\noaffiliation

\date{\today}

\begin{abstract}
We report specific heat measurements on a high quality single crystal of the heavy-fermion compound CePt$_2$In$_7$ in magnetic fields up to 27 T. The zero-field specific heat data above the N\'{e}el temperature, $T_N$, suggest a moderately enhanced value of the electronic specific heat coefficient $\gamma = 180 \; \rm{mJ/K^2mol}$. For $T<T_N$, the data at zero applied magnetic field are consistent with the existence of an anisotropic spin-density wave opening a gap on almost entire Fermi surface, suggesting extreme two-dimensional electronic and magnetic structures for CePt$_2$In$_7$. $T_N$ is monotonically suppressed by magnetic field applied along the $c$-axis. When field is applied parallel to the $a$-axis, $T_N$ first increases at low field up to about 10 T and then decreases monotonically at higher field. Magnetic phase diagram based on specific heat measurements suggests that a field-induced quantum critical point is likely to occur slightly below 60 T for both principal orientations of the magnetic field.
\end{abstract}

% insert suggested PACS numbers in braces on next line
% insert suggested keywords - APS authors don't need to do this
%\keywords{}

%\maketitle must follow title, authors, abstract, \pacs, and \keywords
\maketitle

% body of paper here - Use proper section commands

\section{INTRODUCTION}

Heavy fermion materials are now widely recognized as an ideal playground for studying unusual physics that develops around quantum critical points, i. e. zero-temperature second-order phase transitions. The unusual physics includes unconventional superconductivity~\cite{Pfleiderer2009} and non-Fermi-liquid behavior~\cite{Stewart2001,Stewart2006}. This was indeed observed in most of antiferromagnetic Ce-based heavy-fermion compounds tuned to a quantum critical point by application of pressure~\cite{Settai2007}. A different clean way to tune a heavy-fermion compound to a quantum critical point is by applying magnetic field. Like pressure, field suppression of antiferromagnetism does not induce any disorder contrary to chemical doping. Another advantage of magnetic field as a tuning parameter is that it can be varied continuously across a quantum critical point. However, very high magnetic fields, much higher than available in most laboratories, are sometimes required even when the zero-field N\'{e}el temperature is relatively low. For instance, in cubic CeIn$_3$ with $T_N \approx$ 10 K magnetic field of 61 T is required to completely suppress the antiferromagnetic order~\cite{Ebihara2004}. In CeRhIn$_5$ with tetragonal crystal structure, the zero-field $T_N =$ 3.8 K can be suppressed to zero by application of magnetic field of about 50 T along either [100] or [001] crystallographic directions~\cite{Jiao2015}.

CePt$_2$In$_7$ is a recently discovered heavy fermion antiferromagnet with a N\'{e}el temperature $T_N =$ 5.4 K~\cite{Bauer2010a}. It crystalizes into a body-centered tetragonal crystal structure with space group $P4/mmm$~\cite{Klimczuk2014}. It belongs to the larger family of Ce$_nT_m$In$_{3n+2m}$ ($T$: transition metal, $n =$ 1, 2, $m =$ 0, 1, 2) systems, containing a sequence of $n$ CeIn$_3$ layers intercalated by $m$ $T$In$_2$ layers along the $c$ axis. While cubic CeIn$_3$ ($n =1$, $m =0$) is a completely isotropic system, the layered structures with $m \neq$ 0 lead to strongly anisotropic properties and quasi-two-dimensional Fermi surfaces. Indeed, quasi-two-dimensional Fermi surface sheets were observed in both monolayer ($n = 1$, $m =1$) systems CeCoIn$_5$~\cite{Settai2001,Hall2001}, CeIrIn$_5$~\cite{Haga2001}, CeRhIn$_5$~\cite{Shishido2002,Hall2002} and bilayer ($n = 2$, $m =1$) compounds Ce$_2$RhIn$_8$~\cite{Ueda2004,Jiang2015} and Ce$_2$PdIn$_8$~\cite{Goetze2015}. The degree of two dimensionality is expected to increase with increasing the distance between CeIn$_3$ layers. From this point of view, CePt$_2$In$_7$, in which CeIn$_3$ layers are separated by two PtIn$_2$ layers, is believed to be the most two-dimensional among the so far discovered compounds of the family.

The application of pressure progressively reduces the N\'{e}el temperature in CePt$_2$In$_7$ and finally gives rise to a quantum critical point at $P_c =$ 3.2--3.5 GPa~\cite{Bauer2010a,Sidorov2013,Kurahashi2015}. In polycrystalline samples, superconductivity was observed over a broad pressure range around $P_c$~\cite{Bauer2010a}, while in single crystals superconductivity appears in a much narrower pressure window~\cite{Sidorov2013,Kurahashi2015}. The superconducting critical temperature reaches its highest value $T_c =$ 2.1 K at $P_c$~\cite{Sidorov2013}. This value is one of the highest among Ce-based heavy fermion materials.

Previous measurements of the electrical resistivity on single crystals~\cite{Tobash2012} and of specific heat on polycrystalline samples~\cite{Bauer2010} of CePt$_2$In$_7$ in magnetic field suggest that $T_N$ does not vary with field up to 9 T. Contactless conductivity measurements using a tunnel-diode oscillator circuit in pulsed magnetic fields revealed a clear anomaly at about 45 T~\cite{Altarawneh2011}. The anomaly is temperature-independent and was observed both below and above $T_N$. Furthermore, the anomaly is only weakly anisotropic with respect to the orientation of the magnetic field. While the origin of this feature is not clear, it is certainly not related to the antiferromagnetic transition.

In this paper, we report specific heat measurements on a high quality single crystal of CePt$_2$In$_7$ in high magnetic field. We find that when the magnetic field is applied along the crystallographic $c$-axis, $T_N$ decreases monotonically with increasing magnetic field, which is a usual behavior in heavy-fermion compounds. However, when field is applied along the $a$-axis of the crystal structure, $T_N$ first increases up to about 10 T and then starts to decrease in a close resemblance of CeRhIn$_5$~\cite{Jiao2015}. Our results suggest that a field-induced quantum critical point possibly occurs slightly below 60 T for both principal orientations of the magnetic field.

\section{EXPERIMENTAL DETAILS}

High quality single crystal with the dimensions of 0.47$\times$0.42$\times$0.05 mm$^3$ and a mass of 80 $\mu$g used in the present study was grown by In-self flux method. Details of the sample preparation are given elsewhere~\cite{Kurahashi2015}. The platelet surface is perpendicular to the crystallographic $c$-axis. Specific heat measurements were performed by thermal relaxation technique as described in more details elsewhere~\cite{Lortz2007}.

\section{RESULTS AND DISCUSSION}

\begin{figure}[htb]
\includegraphics[width=8cm]{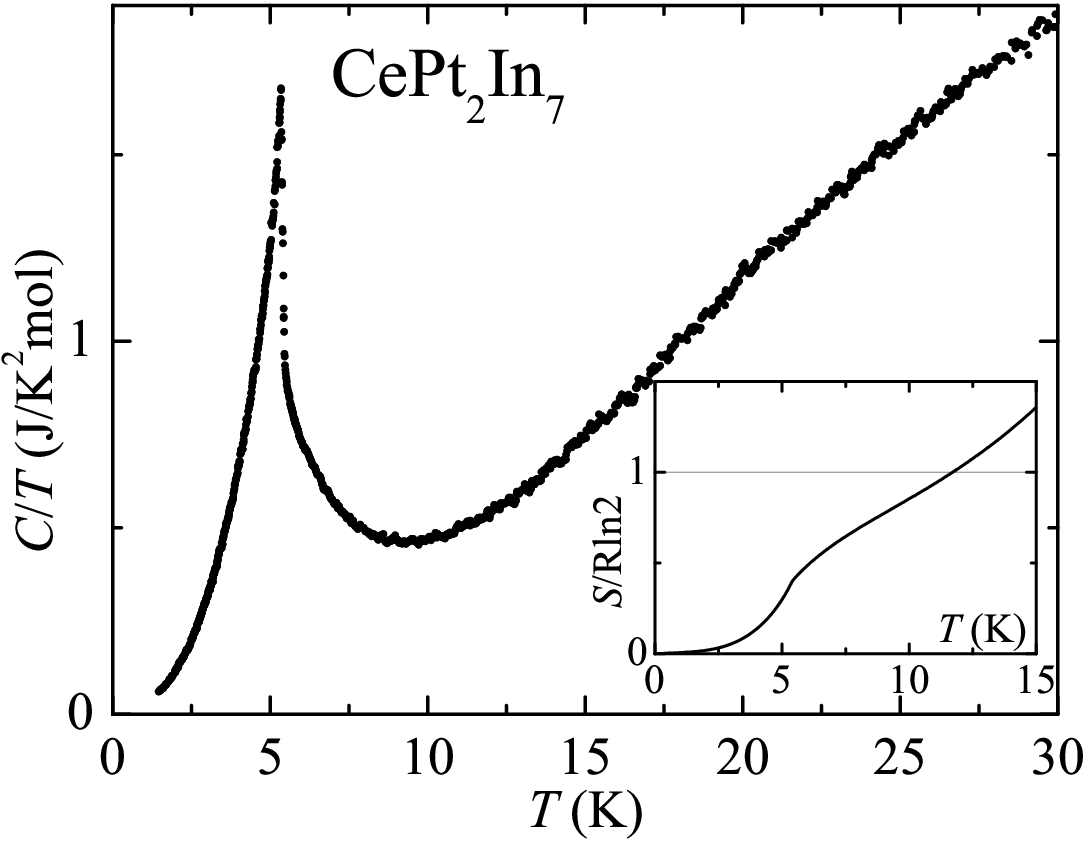}
\caption{\label{fig:Cp0T}Temperature dependence of the specific heat, $C$, divided by temperature, $T$, of CePt$_2$In$_7$ in zero magnetic field. The inset shows the total entropy obtained by integrating $C/T$ versus temperature.}
\end{figure}

Fig.~\ref{fig:Cp0T} shows the zero-field specific heat $C$, divided by temperature, $T$ in CePt$_2$In$_7$. A sharp peak at $T_N$ = 5.4 K corresponds to the onset of antiferromagnetic order in agreement with previous specific heat measurements~\cite{Warren2010,Bauer2010,Bauer2010a,Tobash2012}. However, these measurements were carried out on either polycrystalline samples~\cite{Warren2010,Bauer2010,Bauer2010a} or single crystals contaminated with other phases as evidenced by the presence of specific heat peaks at temperatures other than $T_N$~\cite{Tobash2012}. It is, therefore, not surprising that there is a big discrepancy in the reported zero-field specific heat curves leading to a considerable disagreement in the characteristic parameters extracted from their analysis. Our results, on the contrary, were obtained on a very high quality single crystal, and the absence of any anomalies other than at $T_N$ down to 1.4 K proves that our sample is pure and single-phase.

The total entropy obtained by integrating $C/T$ versus $T$ is shown in the inset of Fig.~\ref{fig:Cp0T}. In Ce-based heavy fermion compounds, magnetic entropy is expected to reach the value of $R \ln2$ at $T_N$. The magnetic entropy is the integral of the magnetic contribution to the specific heat, obtained by subtracting the phonon contribution. This contribution is usually taken as the specific heat of the corresponding non-magnetic La-based material. Unfortunately, this standard approach is not possible in CePt$_2$In$_7$ as LaPt$_2$In$_7$ samples are currently unavailable. Nonetheless, even the total entropy, containing the phonon part, reaches only $0.4R \ln2$ at $T_N$, with the remaining $0.6R \ln2$ recovered at about 12 K. This implies a substantial Kondo-screening of the ordered moments.

\begin{figure}[htb]
\includegraphics[width=8cm]{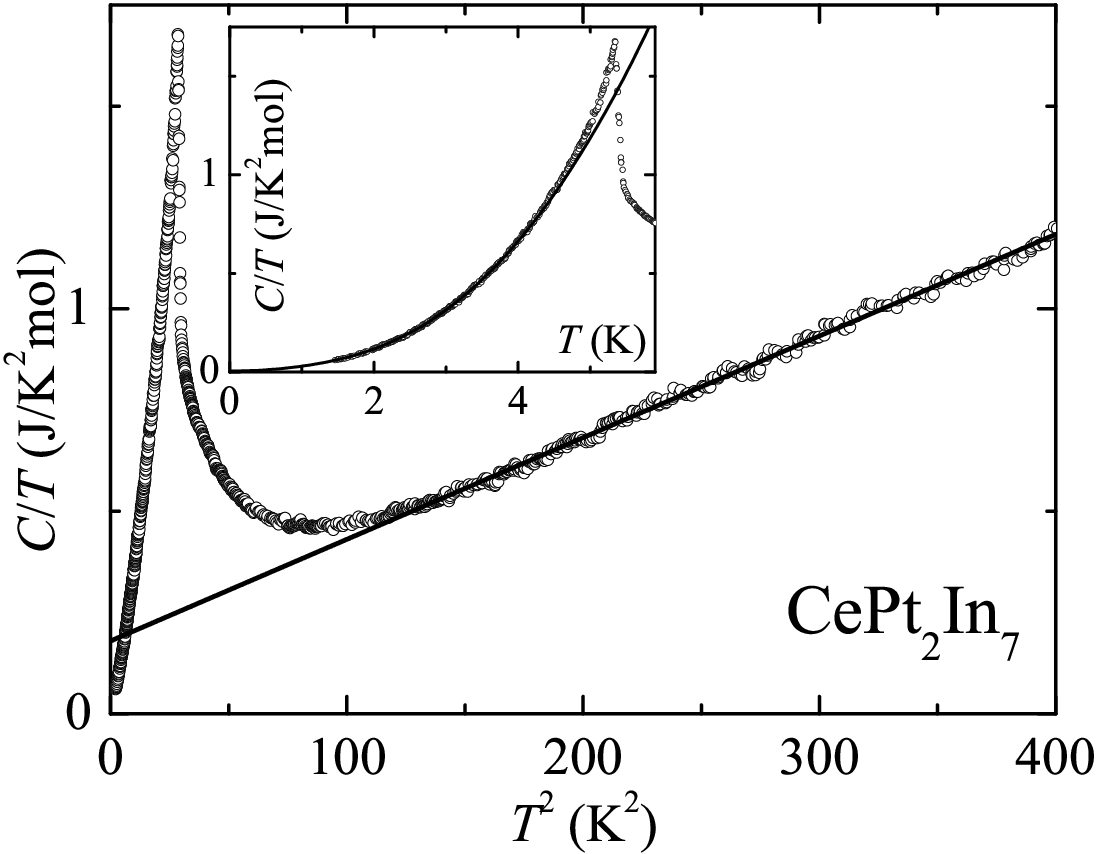}
\caption{\label{fig:Fits0T}The zero-field specific heat, $C$, divided by temperature $T$ vs $T^2$ for CePt$_2$In$_7$. The line is a fit to the standard form of specific heat with electronic and phononic contributions $C/T = \gamma + \beta T^2$. The inset displays a zoom of the data plotted vs $T$ below $T_N$. The line is a fit as described in the text.}
\end{figure}

Fig.~\ref{fig:Fits0T} shows the zero-field specific heat data plotted as a function of $T^2$. The line is a fit to the standard specific heat form $C/T = \gamma + \beta T^2$, where $\gamma$ is the electronic specific heat coefficient and $\beta$ is the lattice Debye term. The fit performed over the temperature range from 11.5 to 20 K yields $\gamma \simeq 180  \; \rm{mJ/K^2mol}$, $\beta \simeq 2.51  \; \rm{mJ/K^4mol}$ with very small standard errors. The lower limit of the chosen temperature range can not be reduced as there is an upturn in $C/T$ between 8 and 11 K, giving rise to a strong deviation of $C/T(T^2)$ from linearity. On the other hand, the choice of the upper limit is rather arbitrary. Increasing this limit by 10\%, to 22 K, yields $\gamma \simeq 197 \; \rm{mJ/K^2mol}$, $\beta \simeq 2.43 \; \rm{mJ/K^4mol}$. Therefore, $\gamma = 180\pm20  \; \rm{mJ/K^2mol}$, and $\beta = 2.5\pm0.1  \; \rm{mJ/K^4mol}$ seem to be a good estimation of the specific heat characteristic parameters.

While the value of $\beta$ found here agrees reasonably well with previously reported results, our value of $\gamma$ is considerably smaller than $340  \; \rm{mJ/K^2mol}$ reported for polycrystalline samples containing a certain amount of impurity phases~\cite{Bauer2010,Bauer2010a}. On the other hand, it is in good agreement with the value extracted from the data obtained on a phase pure polycrystal~\cite{Warren2010}. The only so far reported measurements performed on a single crystal suggest $\gamma = 328  \; \rm{mJ/K^2mol}$~\cite{Tobash2012}. Although this single crystal was contaminated with a small amount of an impurity phase, in this case, the main reason for disagreement with our results seems to be a different temperature range, 8 to 14 K, chosen for the specific heat data analysis. As was already mentioned above, there is an upturn in $C/T$ between 8 and 11 K, suggesting that it would be safer to exclude the data at $T <$ 11 K from the analysis. Indeed, a smaller value of $\gamma$ we find here is more consistent with only moderately enhanced effective masses obtained from de Haas-van Alphen effect measurements~\cite{Altarawneh2011}.

A zoom of the zero-field specific heat data for $T < T_N$ is shown in the inset of Fig.~\ref{fig:Fits0T}. For temperatures below $T_N$, one expects that $C/T$ can be approximated by the sum of an electronic term $\gamma_0$ and an antiferromagntic magnon term $\beta_m T^2$. This simple equation, however, does not fit of our data for $T < T_N$. Indeed, attempts to fit even a small portion of the $C/T(T^2)$ data below $T_N$ by a straight line fail completely yielding unphysical negative values of $\gamma_0$. This sort of behavior has been seen before in other Ce and U compounds where an additional activated term is needed to fit the data~\cite{Bredl1987}. Bredl~\cite{Bredl1987} suggested that the magnetic specific heat data of some heavy fermion materials can be fit using the equation $C_m/T = \gamma_0 + \beta_m T^2 + \beta_{m}^{'} (e^{-\Delta / k_B T}) T^2$ , where the last term is an activated antiferromagnetic magnon term, $\Delta$ being the activated energy. Here the activated antiferromagnetic magnon term arises from an antiferromagnetic spin-density-wave with a gap in the excitation spectrum due to anisotropy. This equation provided a good fit of the CeRhIn$_5$ specific heat data below $T_N$~\cite{Cornelius2000,Cornelius2001}. A fit of our data for $T < 0.8T_N$ by the above equation is shown as a line in the inset of Fig.~\ref{fig:Fits0T}. The fit yields $\gamma_0 = 5\pm4  \; \rm{mJ/K^2mol}$, $\beta_m = 25\pm2  \; \rm{mJ/K^4mol}$, $\beta_{m}^{'} = 78\pm5  \; \rm{mJ/K^4mol}$, and $\Delta / k_B = 6.2\pm0.6 \; \rm{K}$. As we have already mentioned above, the unavailability of the non-magnetic samples of LaPt$_2$In$_7$ does not allow us to subtract the lattice contribution and to obtain pure magnetic specific heat. The above fit was applied to the total specific heat. Therefore, the second term contains also the phonon contribution. This, however, does not effect either the $\gamma_0$ value or the activated antiferromagnetic magnon term. The analysis of the zero-field specific heat results leads us to the conclusion that the magnetically ordered state in CePt$_2$In$_7$ consists of an anisotropic spin-density-wave that opens up a gap of about 6 K on the Fermi surface. A similar conclusion about the nature of the magnetically ordered state was drawn for CeRhIn$_5$~\cite{Cornelius2000,Cornelius2001}, with which CePt$_2$In$_7$ bears a lot of similarities.

From the ratio of the electronic contribution to the specific heat for temperatures below and above $T_N$, we estimate that only a tiny fraction, $\gamma_0 / \gamma \approx 0.03\ (3\%)$, of the Fermi surface remains ungapped below $T_N$. In other words, the Fermi surface is almost entirely gapped in the antiferromagnetic state. For comparison, the ratio of $\gamma_0 / \gamma$ suggests that about 12\% of the Fermi surface remains ungapped below $T_N$ for CeRhIn$_5$~\cite{Cornelius2000,Cornelius2001}, 92\% for less two-dimensional Ce$_2$RhIn$_8$~\cite{Cornelius2000,Cornelius2001}, and 94\% for isotropic CeIn$_3$~\cite{Berton1979}. Reduced dimensionality of the Fermi surface increases the likelihood of a nesting-type magnetic instability. This, in turn, favors the appearance of an anisotropic spin-density-wave that opens up a gap on nested parts of the Fermi surface. Therefore, a greater part of the Fermi surface is expected to be gapped in more two-dimensional materials. From this point of view, as expected, the electronic and magnetic structures in CePt$_2$In$_7$  appear to be the most two-dimensional among the above series.

\begin{figure}[htb]
\includegraphics[width=8cm]{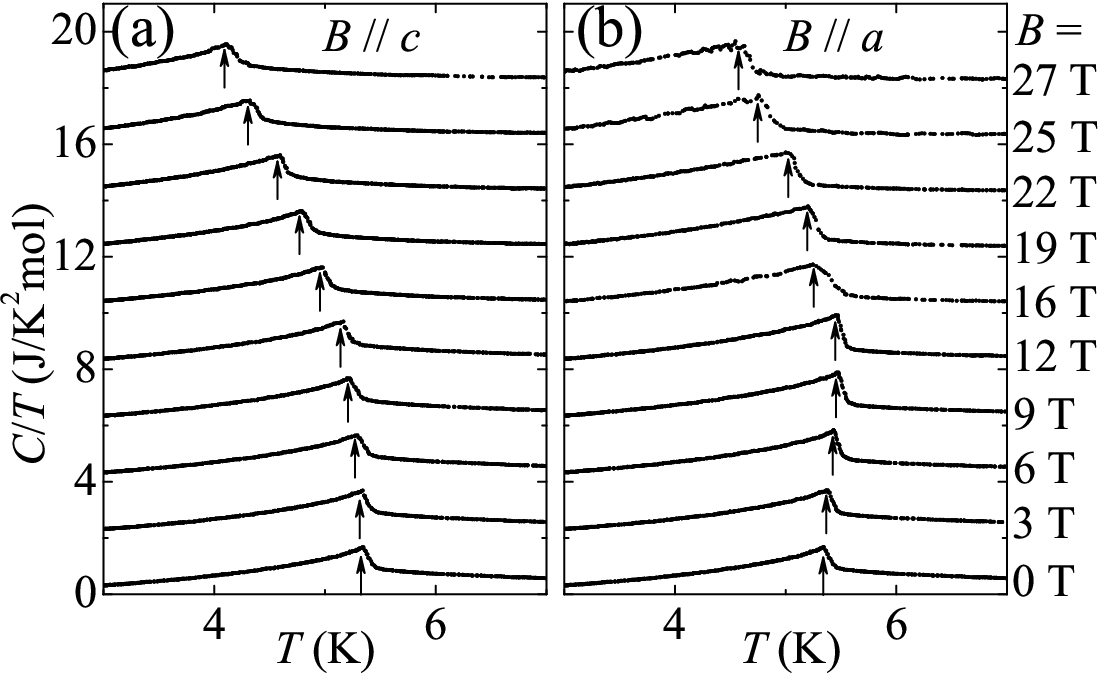}
\caption{\label{fig:Cp_in_field}Specific heat, $C$, divided by temperature, $T$, vs $T$ for CePt$_2$In$_7$ at several values of the magnetic field (shown on the right) applied along the $c$-axis (a) and the $a$-axis (b). Arrows indicate antiferromagnetic ordering temperature $T_N$. The curves are vertically offset by 2 J/K$^2$mol for clarity.}
\end{figure}

Fig.~\ref{fig:Cp_in_field} shows the specific heat divided by temperature in applied magnetic fields for both $B \parallel c$ and $B \parallel a$.  The N\'{e}el temperature is clearly affected by magnetic field applied along either direction.

When the magnetic field is applied along the $c$-axis (Fig.~\ref{fig:Cp_in_field}(a)), $T_N$ decreases monotonically as magnetic field is increased. This is a usual behavior observed in heavy-fermion systems. Here magnetic field suppresses $T_N$ rather slowly; $T_N$ is reduced by only about 20\% at 27 T. A similarly slow decrease of $T_N$ as a function of field along the $c$-axis was observed in CeRhIn$_5$~\cite{Cornelius2001,Kim2001}.

For $B \parallel a$ (Fig.~\ref{fig:Cp_in_field}(b)), the field dependence of $T_N$ shows a very unusual behavior for heavy fermion compounds. Rather than decreasing monotonically with field, $T_N$ first increases by about 2\% up to approximately 10 T before decreasing at higher fields. This unusual behavior was observed in CeRhIn$_5$ for $B \parallel a$ as well, where the rise of $T_N$ is even stronger, about 4\%, at its maximum value also reached at about 10 T~\cite{Cornelius2001,Jiao2015}. In CeRhIn$_5$, this exotic behavior is presumably related to its unusual magnetic structure, an incommensurate helicoidal phase where the magnetic moments form a simple nearest-neighbor antiferromagnet on a square lattice inside the basal plane and spiral transversely along the $c$-axis~\cite{Bao2000,Curro2000,Bao2003,Majumdar2002}. Regarding the magnetic structure of CePt$_2$In$_7$, the existing reports are controversial. Nuclear quadrupolar resonance (NQR) studies performed on polycrystalline samples suggest that antiferromagnetism is commensurate in this material~\cite{Warren2010}. The same conclusion was drawn from positive muon-spin rotation and relaxation measurements also performed on polycrystalline samples~\cite{Mansson2014}. On the contrary, the NQR spectra obtained on single crystals are consistent with the coexistence of incommensurate and commensurate antiferromagnetic components of the magnetic structure~\cite{Sakai2011,Sakai2014}. Clearly, neutron diffraction measurements are needed to determine the exact magnetic structure of CePt$_2$In$_7$ and the degree of its similarity with that of CeRhIn$_5$.

\begin{figure}[htb]
\includegraphics[width=8cm]{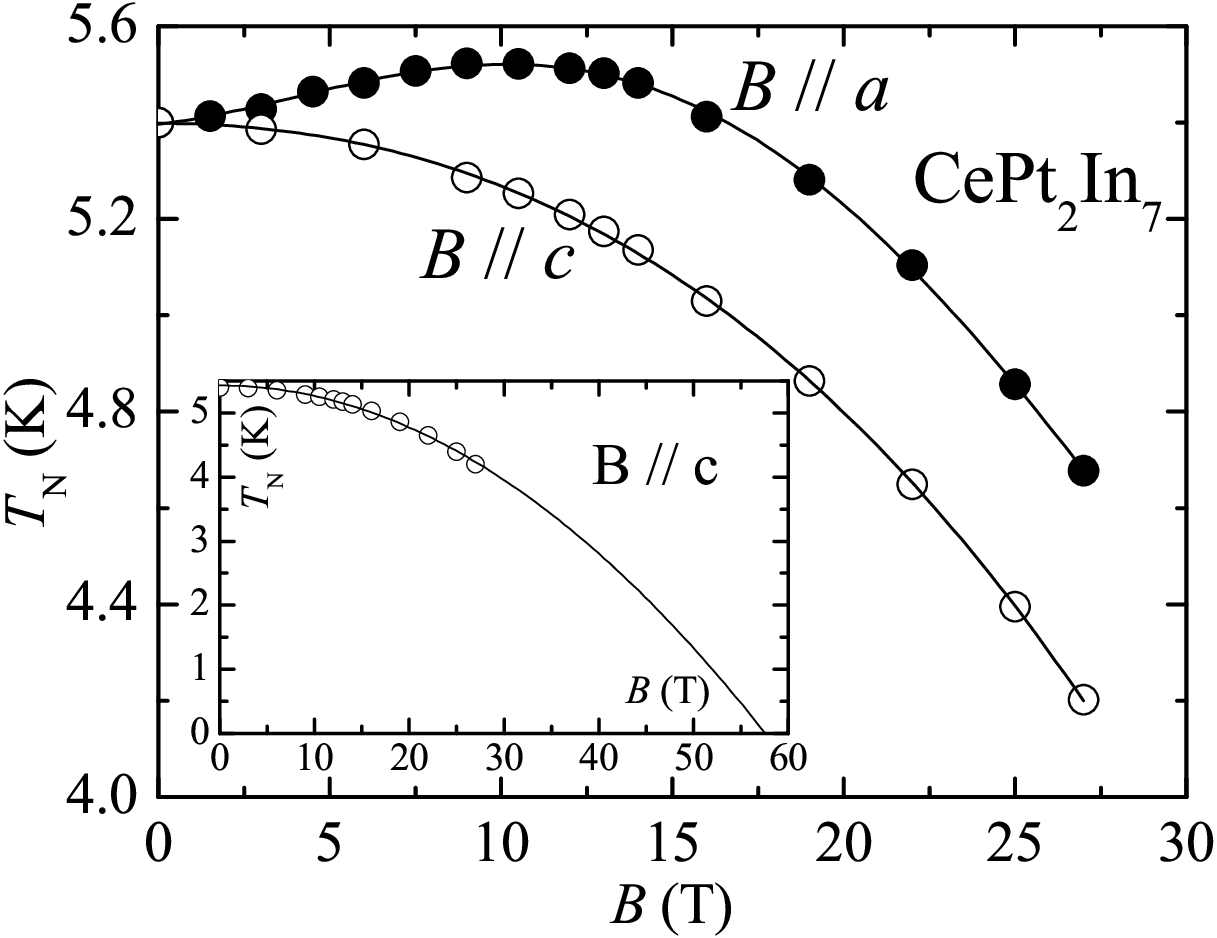}
\caption{\label{fig:Phase_diagram}N\'{e}el temperature as a function of magnetic field applied along the $a$-axis (solid symbols) and the $c$-axis (open symbols). Lines are guides for the eye only. The inset shows field dependence of the N\'{e}el temperature for field applied along the $c$-axis (symbols) together with a scaling function (line) of the form $T_{N0}[1 - (B/B_0)^2]$ with $B_0 =$ 57.6 T.}
\end{figure}

Fig.~\ref{fig:Phase_diagram} shows the resulting phase diagram for both principal orientations of magnetic field. The phase diagram is strikingly similar to that of CeRhIn$_5$, where magnetic field of about 50 T applied along either $a$ or $c$-axis was demonstrated to completely suppress antiferromagnetic order, giving rise to a quantum critical point~\cite{Jiao2015}. As shown in the inset of fig.~\ref{fig:Phase_diagram}, for field applied along the $c$-axis monotonic field dependence of the N\'{e}el temperature is well fit by simple scaling function $T_N(B) = T_{N0}[1 - (B/B_0)^2]$. This function successfully fitted $T_N(B)$ in CePd$_2$Si$_2$~\cite{Sheikin2002} and CeIn$_3$~\cite{Ebihara2004}. In CePt$_2$In$_7$ the fit yields $B_0 = 57.6\pm0.5$ T, the field necessary to suppress $T_N$ to zero when applied along the $c$-axis. Providing the antiferromagnetic transition remains second order down to the lowest temperature, this implies the existence of a field-induced quantum critical point slightly below 60 T for field along the $c$-axis. The value of $B_0$ is very close to the critical field of about 50 T reported for CeRhIn$_5$~\cite{Jiao2015}. Given that $T_N$ decreases monotonically with field applied along the $a$-axis above 10 T and taking into account a close similarity between CePt$_2$In$_7$ and CeRhIn$_5$, it is reasonable to expect a quantum critical point induced by field applied also along the $a$-axis at a value close to $B_0$, i.e. between 50 and 60 T.

\section{SUMMARY}

In summary, we have shown that at zero magnetic field, the specific heat of CePt$_2$In$_7$ below $T_N$ is best described by the sum of an electronic contribution and anisotropic antiferromagnetic spin-density-wave with a gap in the excitation spectrum. This spin-density-wave opens up a gap on almost entire Fermi surface of CePt$_2$In$_7$ emphasizing its two-dimensional electronic and magnetic structures. The antiferromagnetic transition shifts monotonically to lower temperatures in magnetic fields applied along the $c$-axis. When magnetic field is applied along the $a$-axis, the N\'{e}el temperature first increases up to a maximum at 10 T, and then decreases monotonically above that field. Analysis of the field dependence of $T_N$ suggests the suppression of the antiferromagnetic ground state slightly below 60 T, raising the possibility of the existence of a quantum critical point at high field. This prediction calls for further experiments at higher, pulsed magnetic fields.

% If you have acknowledgments, this puts in the proper section head.
\begin{acknowledgments}
We acknowledge the support of the LNCMI-CNRS, member of the European Magnetic Field Laboratory (EMFL).
\end{acknowledgments}

% Create the reference section using BibTeX:
\bibliography{CePt2In7_Phase_Diag}

\end{document}